\documentclass[12pt]{article}
\usepackage{amssymb}
\newcommand{\nnn}[1]{(\ref{#1})}
\newcommand{\operatorname}[1]{{\rm #1}}
\newcommand{\dfrac}[2]{\displaystyle{\frac{#1}{#2}}}
\newcommand{\tfrac}[2]{\textstyle{\frac{#1}{#2}}}

\newenvironment{Proof}[1]{\begin{trivlist} \item[] {\em #1}. }%
{\hfill $///$ \end{trivlist}}

\def\quitelongrightarrow{\relbar\joinrel\relbar\joinrel
                        \relbar\joinrel\relbar\joinrel
                        \relbar\joinrel\relbar\joinrel
                        \relbar\joinrel\relbar\joinrel
                        \relbar\joinrel\relbar\joinrel\longrightarrow}
\let\a=\alpha
\let\b=\beta

\let\d=\delta
\let\D=\Delta
\let\e=\varepsilon

\let\f=\varphi
\let\g=\gamma

\let\h=\ell

\let\k=\kappa
\let\l=\lambda
\let\L=\Lambda
\let\m=\mu

\let\N=\nabla

\let\P=\Pi
\let\r=\rho
\let\s=\sigma
\let\S=\Sigma
\let\t=\tau

\let\y=\psi

\def\bbN{{\mathbb N}}
\def\bbR{{\mathbb R}}

\def\bbZ{{\mathbb Z}}

\def\cC{{\cal C}}

\def\cM{{\cal M}}

\def\cR{{\cal R}}
\def\cS{{\cal S}}

\def\cV{{\cal V}}

\def\gene{\stackrel{{\rm gen}}{=}}
\def\lred{\lambda^{\rm red}}
\def\tl{\tilde\lambda}

\def\gso{{\mathfrak{so}}}

\let\os=\oplus
\let\ot=\otimes
\let\bop=\bigoplus

\def\hom{\operatorname{Hom}}

\def\sgn{\operatorname{sgn}}

\def\spin{\operatorname{Spin}}

\def\tf{\tfrac12}

\def\ath{a^{\underline{{\rm{th}}}}}

\def\circe{{\scriptstyle{\circ}}}

\def\skemastrut{\vrule height 0.4 cm depth 0.2 cm width 0 cm}
\def\skema#1\endskema{\vbox{\def\\{\skemastrut\cr\noalign{\hrule}}%
   \offinterlineskip
   \halign{\vrule\hskip1em
  ##\hskip1em\hfil\vrule&&\hfil\quad$##$\quad\hfil\vrule\cr
\noalign{\hrule}
  #1\skemastrut\crcr\noalign{\hrule}}}}

\newtheorem{Theorem}{Theorem}

\title{Curvature actions on Spin$(n)$ bundles}
\author{Collin Bennett and Thomas Branson}
\date{}

\begin{document}

\maketitle
\abstract{
We compute the number of linearly independent ways
in which a tensor of Weyl type may act upon 
a given irreducible tensor-spinor bundle $\cV$ over a Riemannian manifold.
Together with the analogous but easier problem involving
actions of tensors of Einstein type, this enumerates the
possible curvature actions on $\cV$.} 

\

\section{Introduction}

Let $\cV$ be an irreducible 
$\spin(n)$ vector bundle over an $n$-dimensional Riemannian 
spin manifold $(M,g)$.  The main point of this paper is to give a
simple formula for the number of ways in which the different parts
of the Riemann curvature can {\it act upon} sections of $V$ in an
equivariant way.  

Irreducible $\spin(n)$ bundles are in one-to-one correspondence
with irreducible finite dimensional representations of $\spin(n)$.
These are identical with the irreducible finite dimensional representations
of the Lie algebra $\gso(n)$.  The correspondence of bundles with
representations, or, in the general parlance, {\em modules}, is 
given by the associated bundle construction (see, e.g., \cite{KN}).
Let $\cS$ be the $\spin(n)$-principal 
bundle of spin frames over the base manifold $M$; given a $\spin(n)$-module
$(\f,V)$, we form
the bundle $\cV=\cS\times_\l V$.

The Riemann curvature is a section of a bundle associated to a 
direct sum of irreducible bundles.  These irreducible summands hold
the various parts of the Riemann tensor -- the Weyl conformal curvature
tensor (or its self-dual and anti-self-dual parts in dimension $4$), the
Einstein (trace-free Ricci) 
tensor, and the scalar curvature.  {\em Weitzenb\"ock formulas},
informally speaking, express the difference between two second-order
differential operators on a bundle $\cV$ as a curvature {\em action};
that is, as
a $\spin(n)$-equivariant bundle homomorphism from $\cR\ot\cV$ to 
$\cV$, where $\cR$ is the appropriate bundle of curvature tensors.
Because of the importance of such formulas in geometric analysis, it
is important to understand curvature actions.
For example, the bundle of trace-free symmetric $2$-tensors,
of which the Einstein curvature is a section, acts on the bundle of 
one-forms by (in abstract index notation)
$$
\f_a\mapsto\s_a{}^b\f_b\,.
$$
Algebraic Weyl tensors, i.e.\ sections of the bundle in which the Weyl
curvature lives, are capable of acting on two-forms $\f_{ab}$ via
$$
\f_{ab}\mapsto C_{ab}{}^{cd}\f_{cd}\,.
$$

We find the number of actions of each part of the Riemann curvature 
tensor (or, more precisely, each part of a generic algebraic Riemann
curvature tensor) on each irreducible bundle by computing the dimension
of the space of equivariant homomorphisms as described above.  By
the associated
bundle construction and the fact that the various parts of the 
curvature are sections of specific $\spin(n)$-bundles, 
this is the same problem as that of computing
the dimension of $\hom_{\gso(n)}(U\ot V,V)$ for 
certain irreducible $\gso(n)$-modules $U$, and a general irreducible
$\gso(n)$-module $V$.  The problem is the same since the associated 
bundle construction 
``promotes'' each module homomorphism to the bundle setting;  
bundle homomorphisms ``demote'' to module homomorphisms just by 
evaluation at any point.

\section{Some facts from representation theory}

We refer to \cite{Hu} and \cite{He} 
for the following standard facts on $\gso(n)$.
Let $n\ge 2$.  
{\em Integral weights} for $\gso(n)$
are $\h$-tuples, $\h=[n/2]$, consisting entirely
of integers, or entirely of half-integers:
$$
\Pi=\bbZ^\h\cup(\tfrac12+\bbZ)^\h.
$$
For each finite-dimensional representation $(\f,V)$ of $\gso(n)$,
the vector space $V$ is a direct sum of {\em weight spaces} $V_\m\,$, for
various $\m$ in $\Pi$.  We put
$$
\Pi(\f)=\{\m\in\P\mid V_\m\ne 0\}.
$$
The direct sum decomposition 
$$
V=\bop_{\m\in\P(\f)}V_\m
$$
is of course not $\gso(n)$-invariant in general; each weight space
is, however, invariant under a maximal abelian subalgebra of $\gso(n)$.
The {\em multiplicity} of a weight $\m$ in a finite dimensional representation
$(\f,V)$ is the dimension of $V_\m\,$.

A {\em dominant integral weight} is an integral weight $\l$ which satisfies 
the inequality condition
\begin{equation}\label{dom}
\begin{array}{ll}
\l_1\ge\l_2\ge\cdots\ge\l_{\h-1}\ge|\l_\h|\,, & n\ {\rm even}, \\
\l_1\ge\l_2\ge\cdots\ge\l_{\h-1}\ge \l_\h\ge 0,\qquad & n\ {\rm odd}.
\end{array}
\end{equation}
The finite-dimensional irreducible representations of $\gso(n)$ are
parameterized by dominant integral weights, and in fact, the
dominant weight parameter $\l$ associated to a given irreducible
representation is its (lexicographically) highest weight.
In what follows, we fix a dominant integral weight $\l$, and study
the corresponding irreducible representation $V(\l)$.
The set of weights of $V(\l)$ will be denoted $\P(\l)$, and
the multiplicity of the weight $\m$ in the representation $V(\l)$
will be denoted by $m_\l(\m)$.

In writing dominant weights, we shall sometimes use the convention
of Strichartz, in which terminal strings of zeros are omitted from
the notation.  Thus, for example, we write $(1,1,0,\ldots,0)$ as
$(1,1)$.

Let $e_a$ be the $\h$-tuple with $1$ in the $\ath$ entry, and $0$
in all other entries.  Let
$$
\a_{ab}^\pm=e_a\pm e_b\,.
$$
The {\em roots} of
$\gso(n)$ are the weights of the adjoint representation.
With their multiplicities, these are
$$
\skema
root $\m$ & {\rm multiplicity} \\
$0$   & \h \\
$e_a-e_b\ (a\ne b)$ & 1 \\
$e_a+e_b\ (a\ne b)$ & 1 \\                          
$-e_a-e_b\ (a\ne b)$ & 1 \\
$e_a$ & n-2\h \\  
$-e_a$ & n-2\h
\endskema
$$
The sum of the (lexicographically) positive roots in $\gso(n)$ is
$$
2\r=(n-2,n-4,\ldots,n-2\h)=\left\{\begin{array}{l}
(n-2,n-4,\ldots,2,0),\ n\ {\rm even}, \\
(n-2,n-4,\ldots,3,1),\ n\ {\rm odd}.
\end{array}\right.
$$

Let $\m\in\P$, and let $\D^+$ be the set of positive roots.  One knows that
that a {\em dominant} weight $\m$ lies in $\P(\l)$ if and only if
\begin{equation}\label{subord}
\m+\sum_{\a\in\D^+}k_\a\a=\l
\end{equation}
for some list of natural numbers $k_\a\,$.  If \nnn{subord} holds,
$\m$ is said to be {\em subordinate} to $\l$.
{\em Freudenthal's formula} allows one to inductively
compute the multiplicities
$m_\l(\m)$, for $\m\in\P(\l)$, starting with the fact that $m_\l(\l)=1$:
\begin{equation}\label{freu}
(\|\tilde\l\|^2-\|\tilde\m\|^2)m_\l(\m)=2\sum_{\a\in\D^+}
\sum_{j=1}^\infty m_\l(\m+j\a)(\m+j\a,\a).
\end{equation}
Here if $\m$ is a weight, then $\tilde\m=\m+\r$.  The 
inner product $(\cdot,\cdot)$ on the right is the standard inner product
in $\bbR^\h$.  A very useful fact, which we shall often use implicitly,
is that
if $\l$ is dominant, then $\tilde\l$ is {\em strictly dominant} --
it satisfies a version of \nnn{dom} in which the $\ge$ signs are replaced
by $>$ signs.
Note that the sum on the right in \nnn{freu}
is finite because $\P(\l)$ is finite.  Freudenthal's
formula expresses the multiplicity $m_\l(\m)$
in terms of multiplicities $m_\l(\nu)$ for $\nu$ lexicographically higher
than $\m$; thus the computation is truly inductive.  
To make the computation run, we simply have to find all
positive root strings through dominant weights in $\P(\l)$.
Note that if $\m$ is dominant and subordinate to $\l$, then
$\m+j\a$ ($j\in\bbN$, $\a\in\D^+$) is dominant, and is either subordinate
to $\l$, or outside of $\P(\l)$.

To get the multiplicities of non-dominant weights in $V(\l)$, one
uses the action of the {\em Weyl group} $W$.  In the case of $\gso(n)$,
the Weyl group acts on $\P$ as follows.
If $n$ is odd, a given $w\in W$ acts
by permutation, together with any number of sign changes on entries 
of a weight.  If $n$ is even, a given $w\in W$ acts by 
permutation and an even number of 
sign changes.  
An important point is that $W$ acts on $\P(\l)$ (not just on $\P$),
and that 
\begin{equation}\label{WeylPl}
m_\l(w\cdot\m)=m_\l(\m),\qquad{\rm all}\ \m\in\P(\l),\ w\in W.
\end{equation}
Since each Weyl group orbit clearly contains a unique dominant weight,
Freudenthal's formula solves the problem of finding {\em all} weights,
with their multiplicities, in $V(\l)$.

Given two dominant integral weights $\s$ and $\l$, an important and
much-studied problem is that of computing the direct sum decomposition
of $V(\s)\ot V(\l)$.  By Weyl's Theorem \cite{Hu}, any finite-dimensional
representation of $\gso(n)$ is completely reducible ($\gso(n)$ being
semisimple):
$$
V(\s)\ot V(\l)\cong_{\gso(n)}\bop_{\k\in\P^{{\rm DI}}}
\cM_\k(V(\s)\ot V(\l)),
$$
where $\Pi^{{\rm DI}}$ is the set of dominant integral weights.
Here the multiplicities $\cM_\k(V(\s)\ot V(\l))$, which give the 
number of isomorphic copies of $V(\k)$ in 
$V(\s)\ot V(\l)$, are natural numbers; by finite-dimensionality,
all but a finite number of these vanish.
The {\em Brauer-Kostant formula} \cite{brauer,kostant} expresses the numbers
$\cM_\k(V(\s)\ot V(\l))$ in terms of $\l$ and all the weights, with
multiplicities, of $V(\s)$.  This is ideal for the type of problem
we have here, in which we would like to compute information about 
the tensor product of a fixed representation (that to which algebraic
Weyl tensors are associated) with an arbitrary representation -- we
just need the weights and multiplicities of the fixed representation.

To state the Brauer-Kostant formula, we need the action of the Weyl
group $W$ of $\gso(n)$ on $\P$.  
The {\em sign} of a Weyl group element, $\sgn\,w$, is
the sign of the permutation times the number of sign changes.
The Brauer-Kostant formula says that
$$
\cM_\l(V(\s)\ot V(\t))=\sum_{\m\in\Pi}m_\s(\m)\sum_{w\in W}
(\sgn\,w)\d_{w\cdot\tilde\l}^{\tilde\t-\m}\,,
$$
where $\d$ is the Kronecker delta.

\section{Computations}\label{compusection}

By \cite{Str}, the bundle of algebraic Weyl tensors is associated to
the representation $V(2,2)$ for $n\ge 5$, and to $V(2,2)\os V(2,-2)$
for $n=4$.  Thus the first task is to find the weights, with
multiplicities, of these modules.  As noted above, each Weyl group
orbit contains a unique dominant weight, so we need only list the
multiplicities of dominant weights subordinate to $(2,2)$ (and to $(2,-2)$
when $n=4$).  In the following theorem, we get these weights and
multiplicities, as well as the sizes of the Weyl group orbits
associated to each.  This latter bit of information is not strictly
necessary to our calculations, but allows us to accomplish a
reassuring check: that the weight multiplicities add up to the
dimension of the module.  In addition, we give the contribution of
each Weyl group orbit to the expression on the right in the
Brauer-Kostant formula for $\cM_\l(V(2,2)\ot V(\l))$ (and, when $n=4$,
the same expression with $(2,-2)$ in place of $(2,2)$).
Multiplying each contribution by the corresponding 
multiplicity and adding gives
the number $\cM_\l(V(2,2)\ot V(\l))$ (and the same with $(2,-2)$ in 
place of $(2,2)$ when $n=4$); this information is collected in
Theorems \ref{ThmEven}, \ref{ThmFour}, and \ref{ThmOdd} below.

To write the contributions to the right-hand side of the Brauer-Kostant
formula, we need to define some parameters based on the dominant
weight $\l$.  If $1\le k\le\h$, let
$$
\e_{\k_{\h-k}\,,\ldots,\k_\h}=\left\{
\begin{array}{ll}
1,\ & (\l_{\h-k}\,,\ldots,\l_\h)=(\k_{\h-k}\,,\ldots,\k_\h), \\
0 & {\rm otherwise},
\end{array}\right.
$$
and let
$$
\l^{(k)}=(\l_1\,,\ldots,\l_{\h-k})
$$
Furthermore, let
$$
\begin{array}{rl}
T&=\#\{a\mid\l_a=\l_{a+1}=\l_{a+2}\}, \\
D&=\#\{a\mid\l_{a+1}=\l_a-1\},        \\
P&=\#\{a\mid\l_{a+1}=\l_a\},          \\ 
S&=\#\{\{a,b\}\mid\l_{a+1}=\l_a\,,\,\l_{b+1}=\l_b\,,\,
\#\{a,a+1,b,b+1\}=4\}.
\end{array}
$$
To paraphrase, $T$ is the number of ``flat triples'',
$D$ is the number of ``1-drops'', $P$ the number of ``flat pairs'',
and $S$ is the number of disjoint pairs of
flat pairs.

\begin{Theorem}\label{WeightTables} 
The dominant weights of $V(2,2)$, with their multiplicities, 
orbit sizes, and orbit contributions to 
$\cM_\l(V(2,2)\ot V(\l))$ (assuming, when $n$ is even, that $\l_\h\ge 0$)
are given by the following
tables:

\vspace{0.2in}

$n\ge 10$ even:
\smallskip
$$
\skema
{\rm weight} $\m$ & {\rm multiplicity} & {\rm orbit\ size}\ \#(W\cdot\m) 
& {\rm contribution} \\
$(2,2)$   & 1         & 2\h(\h-1)                   & -T-D-\e_{0,0,0}
                                                         -\e_{1/2,1/2}       \\
$(2,1,1)$ & 1         & 4\h(\h-1)(\h-2)             & 2T+2\e_{0,0,0}  
                                                        -\e_{0,0}P(\l^{(2)}) \\
$(2)$     & \h-2      & 2\h                         & \e_{0,0}               \\
$(1,1,1,1)$ & 2       & \frac23\h(\h-1)(\h-2)(\h-3) & S+\e_{0,0}P(\l^{(2)})  \\
$(1,1)$   & 2\h-3     & 2\h(\h-1)                   & -P-\e_{0,0}            \\
$0$       & \h(\h-1)  & 1                           & 1
\endskema
$$
\medskip
$n=8$:
\smallskip
$$
\skema
{\rm weight} $\m$ & {\rm multiplicity} & {\rm orbit\ size}\ \#(W\cdot\m) 
& {\rm contribution} \\
$(2,2)$   & 1         & 2\h(\h-1)=24                & -T-D-\e_{0,0,0}
                                                         -\e_{1/2,1/2}       \\
$(2,1,1)$ & 1         & 4\h(\h-1)(\h-2)=96          & 2T+2\e_{0,0,0}  
                                                        -\e_{0,0}P(\l^{(2)}) \\
$(2)$     & \h-2=2    & 2\h=8                       & \e_{0,0}               \\
$(1,1,1,1)$ & 2       & 8                           & S                      \\
$(1,1,1,-1)$ & 2      & 8                           & \e_{0,0}P(\l^{(2)})    \\
$(1,1)$   & 2\h-3=5   & 2\h(\h-1)=24                & -P-\e_{0,0}            \\
$0$       & \h(\h-1)=12  & 1                        & 1
\endskema
$$
\medskip
$n=6$:
\smallskip
$$
\skema
{\rm weight} $\m$ & {\rm multiplicity} & {\rm orbit\ size}\ \#(W\cdot\m) 
& {\rm contribution} \\
$(2,2)$   & 1         & 2\h(\h-1)=12                & -T-D-\e_{0,0,0}
                                                         -\e_{1/2,1/2}       \\
$(2,1,1)$ & 1         & 12                          & T+\e_{0,0,0}           \\
$(2,1,-1)$ & 1        & 12                          & T+\e_{0,0,0}           \\
$(2)$     & \h-2=1    & 2\h=6                       & \e_{0,0}               \\
$(1,1)$   & 2\h-3=3   & 2\h(\h-1)=12                & -P-\e_{0,0}            \\
$0$       & \h(\h-1)=6  & 1                         & 1
\endskema
$$
\medskip
$n\ge 9$ odd:
\smallskip
$$
\skema
{\rm weight} $\m$ & {\rm multiplicity} & {\rm orbit\ size}\ \#(W\cdot\m) 
& {\rm contribution} \\
$(2,2)$   & 1            & 2\h(\h-1)                   & -T-D-\e_{0,0}       \\
$(2,1,1)$ & 1            & 4\h(\h-1)(\h-2)         & 2T+\e_{1/2}P(\l^{(1)})  \\
$(2,1)$   & 1            & 4\h(\h-1)                   & 2\e_{0,0}           \\
$(2)$     & \h-1         & 2\h                         & -\e_{1/2}           \\
$(1,1,1,1)$ & 2          & \frac23\h(\h-1)(\h-2)(\h-3) & S                   \\
$(1,1,1)$ & 2            & \frac43\h(\h-1)(\h-2)       & \e_0P(\l^{(1)})     \\
$(1,1)$   & 2(\h-1)      & 2\h(\h-1)                   & -P                  \\
$(1)$     & 2(\h-1)      & 2\h                         & -\e_0               \\
$0$       & (\h+1)(\h-1) & 1                           & 1
\endskema
$$
\medskip
$n=7$:
\smallskip
$$
\skema
{\rm weight} $\m$ & {\rm multiplicity} & {\rm orbit\ size}\ \#(W\cdot\m) 
& {\rm contribution} \\
$(2,2)$   & 1            & 2\h(\h-1)=12                & -T-D-\e_{0,0}       \\
$(2,1,1)$ & 1            & 4\h(\h-1)(\h-2)=24      & 2T+\e_{1/2}P(\l^{(1)})  \\
$(2,1)$   & 1            & 4\h(\h-1)=24                & 2\e_{0,0}           \\
$(2)$     & \h-1=2       & 2\h=6                       & -\e_{1/2}           \\
$(1,1,1)$ & 2            & \frac43\h(\h-1)(\h-2)=8     & \e_0P(\l^{(1)})     \\
$(1,1)$   & 2(\h-1)=4    & 2\h(\h-1)=12                & -P                  \\
$(1)$     & 2(\h-1)=4    & 2\h=6                       & -\e_0               \\
$0$       & (\h+1)(\h-1)=8 & 1                         & 1
\endskema
$$
\medskip
$n=5$:
\smallskip
$$
\skema
{\rm weight} $\m$ & {\rm multiplicity} & {\rm orbit\ size}\ \#(W\cdot\m) 
& {\rm contribution} \\
$(2,2)$   & 1            & 2\h(\h-1)=4                 & -D-\e_{0,0}         \\
$(2,1)$   & 1            & 4\h(\h-1)=8                 & 2\e_{0,0}           \\
$(2)$     & \h-1=1       & 2\h=4                       & -\e_{1/2}           \\
$(1,1)$   & 2(\h-1)=2    & 2\h(\h-1)=4                 & -P                  \\
$(1)$     & 2(\h-1)=2    & 2\h=4                       & -\e_0               \\
$0$       & (\h+1)(\h-1)=3 & 1                         & 1
\endskema
$$
\medskip
$n=4$, module $V(2,2)$:
\smallskip
$$
\skema
{\rm weight} $\m$ & {\rm multiplicity} & {\rm orbit\ size}\ \#(W\cdot\m) 
& {\rm contribution} \\
$(2,2)$   & 1         & 2                           & -\e_{1/2,1/2}          \\
$(1,1)$   & 1         & 2                           & -\e_{0,0}              \\
$0$       & 1         & 1                           & 1
\endskema
$$
\medskip
$n=4$, module $V(2,-2)$:
\smallskip
$$
\skema
{\rm weight} $\m$ & {\rm multiplicity} & {\rm orbit\ size}\ \#(W\cdot\m) 
& {\rm contribution} \\
$(2,-2)$   & 1         & 2                & -D \\
$(1,-1)$   & 1         & 2                & -P \\
$0$        & 1         & 1                & 1
\endskema
$$
\medskip
(In the very last table, the ``contribution'' is to
$\cM_\l(V(2,-2)\ot V(\l))$ rather than $\cM_\l(V(2,2)\ot V(\l))$.) 
\end{Theorem}

\begin{Proof}{Proof}
The identification of the weights is a straightforward 
application of \nnn{subord}, and the orbit sizes
are immediate from the above description of the Weyl group.
The multiplicities 
may be computed inductively, as outlined above, once all positive
root strings through dominant weights are identified.

If $n\ge 10$ is even, 
the positive root strings through dominant weights are described by
\begin{equation}\label{RtStrE}
\begin{array}{l}
(2,1,1)+\a_{23}^-=(2,2), \\ 
(1,1,1,1)+\a_{ab}^-\sim(2,1,1),\qquad(4\ge b>a), \\
(2)+\a_{ab}^\pm\sim(2,1,1)\qquad(b>a>1), \\
(1,1)+\a_{ab}^\pm\sim(1,1,1,1)\qquad(b>a>2), \\
(1,1)+\a_{1a}^\pm\sim(2,1,1)\qquad(a>2), \\
(1,1)+\a_{2a}^\pm\sim(2,1,1)\qquad(a>2), \\
(1,1)+\a_{12}^-=(2)\qquad(a>2), \\
(1,1)+\a_{12}^+=(2,2), \\
0+\a_{ab}^\pm\sim(1,1)\ {\rm and}\ 0+2\a_{ab}^\pm\sim(2,2)\qquad(b>a).
\end{array}
\end{equation}
Here $\m\sim\nu$ means that $\m$ and $\nu$ are in the same Weyl group
orbit.

If $n\ge 9$ is odd, we have the root strings \nnn{RtStrE}
through weights, as well as the following:
\begin{equation}\label{RtStrO}
\begin{array}{l}
(2,1)+e_2=(2,2), \\
(2,1)+e_a\sim(2,1,1)\qquad(a>2), \\
(1,1,1)+\a_{ab}^-\sim(2,1)\qquad(3\ge b>a), \\
(1,1,1)+e_a\sim(2,1,1)\qquad(3\ge a), \\
(1,1,1)+e_a\sim(1,1,1,1)\qquad(a>3), \\
(2)+e_a\sim(2,1)\ {\rm and}\ (2)+2e_a\sim(2,2)\qquad(a>1), \\
(1,1)+e_a\sim(2,1)\qquad(2\ge a), \\
(1,1)+e_a\sim(1,1,1)\ {\rm and}\ (1,1)+2e_a\sim(2,1,1)\qquad(a>2), \\
(1)+\a_{1a}^\pm\sim(2,1)\qquad(a>1), \\
(1)+\a_{ab}^\pm\sim(1,1,1)\qquad(b>a>1), \\
(1)+e_1=(2), \\
(1)+e_a\sim(1,1)\ {\rm and}\ (1)+2e_a\sim(2,1)\qquad(a>1), \\
0+e_a\sim(1)\ {\rm and}\ 0+2e_a\sim(2)\qquad(b>a).
\end{array}
\end{equation}

The contributions to
$\cM_\l(V(2,2)\ot V(\l))$ for $n\ge 9$
arise as follows.  

In case of a flat triple
$\l_a=\l_{a+1}=\l_{a+2}\,$,
$$
\begin{array}{l}
(\ldots,\tilde\l_a\,,\tilde\l_{a+1}\,,\tilde\l_{a+2}\,,\ldots)
-(\ldots,2,0,-2,\ldots) \\
\qquad=(\ldots,\tilde\l_a\,,\tilde\l_a-1,\tilde\l_a-2,\ldots)
-(\ldots,2,0,-2,\ldots) \\
\qquad=(\ldots,\tilde\l_a-2,\tilde\l_a-1,\tilde\l_a,\ldots) \\
\qquad={\rm(transposition)}\cdot\tilde\l.
\end{array}
$$
In addition, 
\begin{equation}\label{TContrT}
\begin{array}{l}
(\ldots,\tilde\l_a\,,\tilde\l_{a+1}\,,\tilde\l_{a+2}\,,\ldots)
-(\ldots,2,-1,-1,\ldots) \\
{}\ =(\ldots,\tilde\l_a-2,\tilde\l_a\,,\tilde\l_a-1,\ldots)
={\rm(3-cycle)}\cdot\tilde\l.
\end{array}
\end{equation}
and
\begin{equation}\label{TContrH}
\begin{array}{l}
(\ldots,\tilde\l_a\,,\tilde\l_{a+1}\,,\tilde\l_{a+2}\,,\ldots)
-(\ldots,1,1,-2,\ldots) \\
{}\ =(\ldots,\tilde\l_a-1,\tilde\l_a-2\,,\tilde\l_a\,,\ldots)
={\rm(3-cycle)}\cdot\tilde\l.
\end{array}
\end{equation}
This accounts for the $T$ contributions in either dimension parity.

For a $1$-drop $\l_{a+1}=\l_a-1$,
\begin{equation}\label{OneDrop}
\begin{array}{rl}
\l&=(\ldots,\tilde\l_a\,,\tilde\l_{a+1}\,,\ldots)-(\ldots,2,-2,\ldots) \\
&=(\ldots,\tilde\l_a\,,\tilde\l_a-2,\ldots)-(\ldots,2,-2,\ldots) \\
&=(\ldots,\tilde\l_a-2,\tilde\l_a\,,\ldots) \\
&={\rm(transposition)}\cdot\tilde\l.
\end{array}
\end{equation}
This accounts for the $D$ contribution in either dimension parity.

In case of a pair of disjoint pairs, 
$\l_{a+1}=\l_a\,$, $\l_{b+1}=\l_b\,$, $\#\{a,a+1,b,b+1\}=4$, we
may assume $a+1<b$; then
$$
\begin{array}{l}
(\ldots,\tl_a\,,\tl_{a+1}\,,\ldots,\tl_b\,,\tl_{b+1}\,,\ldots)
-(\ldots,1,-1,\ldots,1,-1) \\
\qquad=(\ldots,\tl_a\,,\tl_a-1,\ldots,\tl_b\,,\tl_b-1,\ldots)
-(\ldots,1,-1,\ldots,1,-1) \\
\qquad=(\ldots,\tl_a-1,\tl_a\,,\ldots,\tl_b-1,\tl_b\,,\ldots) \\
\qquad={\rm(product\ of\ 2\ transpositions)}\cdot\tl.
\end{array}
$$
This accounts for the $S$ contribution in either dimension parity.

For each pair $\l_a=\l_{a+1}\,$, 
\begin{equation}\label{Pair}
\begin{array}{l}
(\ldots,\tl_a\,,\tl_{a+1}\,,\ldots)
-(\ldots,1,-1,\ldots) \\
{}\ =(\ldots,\tl_a\,,\tl_a-1,\ldots)
-(\ldots,1,-1,\ldots) \\
{}\ =(\ldots,\tl_a-1,\tl_a\,,\ldots) \\
{}\ ={\rm(transposition)}\cdot\tl.
\end{array}
\end{equation}
This accounts for the $P$ contribution in either dimension parity.

The contribution of 1, coming from the weight 0, is clear.

If $n\ge 6$ is even and $\l_{\h-2}=\l_{\h-1}=\l_\h=0$, then
$$
\tl-(\ldots,2,0,-2)=
(\ldots,2,1,0)-(\ldots,2,0,-2)=(\ldots,0,1,2)
={\rm(transposition)}\cdot\tl.
$$
Furthermore, 
\begin{equation}\label{EoooContr}
\begin{array}{l}
\tl-(\ldots,2,-1,-1)=(\ldots,0,2,1)={\rm(3-cycle)}\cdot\tl, \\
\tl=(\ldots,1,1-2)=(\ldots,1,0,2)={\rm(3-cycle)}\cdot\tl,
\end{array}
\end{equation}
This accounts for the $\e_{0,0,0}$ contributions in the even case.

If $n$ is even and $\l_{\h-1}=\l_\h=1/2$, then
\begin{equation}\label{EHalfHalf}
\begin{array}{rl}
\tl-(\ldots,2,2)
&=(\ldots,\frac32\,,\frac12)-(\ldots,2,2) \\
&=(\ldots,-\frac12\,,-\frac32) \\
&={\rm(transposition\ and\ 2\ sign\ changes)}
\cdot\tl.
\end{array}
\end{equation}
This accounts for the $\e_{1/2,1/2}$ contribution in the even case.

If $n$ is even and $\l_{\h-1}=\l_\h=0$, we have
$$
\tl-(\ldots,2,0)=(\ldots,1,0)-(\ldots,2,0)
=(\ldots,-1,0)={\rm(2\ sign\ changes)}\cdot\tl.
$$
Furthermore, 
\begin{equation}\label{EooH}
\tl-(\ldots,1,1)=(\ldots,0,-1)={\rm(transposition\ and\ 2\ sign\ changes)}
\cdot\tl.
\end{equation}
(Note that the contribution from 
$$
\tl-(\ldots,1,-1)=(\ldots,0,1)=
{\rm(transposition)}\cdot\tl$$ 
has already been counted, and is
included in the $P$ contribution.)
This accounts for the $\e_{0,0}$ contributions in the even case.

There are also some $\e_{0,0}P(\l^{(2)})$ contributions in the even
case; these arise as follows.  If $\l_a=\l_{a+1}\,$, $a+1<\h-1$,
and $\l_{\h-1}=\l_\h=0$, we have
\begin{equation}\label{EooPlHContr}
\begin{array}{l}
\tl-(\ldots,1,-1,\ldots,2,0) \\
{}\ =(\ldots,\tl_a\,,\tl_a-1,\ldots,1,0)-(\ldots,1,-1,\ldots,2,0) \\
{}\ =(\ldots,\tl_a-1,\tl_a\,,\ldots,-1,0) \\
{}\ ={\rm(transposition\ and\ 2\ sign\ changes)}
\cdot\tl,
\end{array}
\end{equation}
and
$$
\begin{array}{rl}
\tl-(\ldots,1,-1,\ldots,1,1)&=
(\ldots,\tl_a\,,\tl_a-1,\ldots,1,0)-(\ldots,1,-1,\ldots,1,1) \\
&=(\ldots,\tl_a-1,\tl_a\,,\ldots,0,-1) \\
&={\rm(2\ transpositions\ and\ 2\ sign\ changes)}
\cdot\tl.
\end{array}
$$

If $n$ is odd and $\l_{\h-1}=\l_\h=0$, then
the calculation \nnn{EHalfHalf} applies.  Furthermore,
\begin{displaymath}
\begin{array}{rl}
\tl-(\ldots,1,2)
&=(\ldots,\frac32\,,\frac12)-(\ldots,1,2) \\
&=(\ldots,\frac12\,,-\frac32) \\
&={\rm(transposition\ and\ 1\ sign\ change)}
\cdot\tl,
\end{array}
\end{displaymath}
and
\begin{displaymath}
\begin{array}{rl}
\tl-(\ldots,1,2)
&=(\ldots,\frac32\,,\frac12)-(\ldots,2,-1) \\
&=(\ldots,-\frac12\,,\frac32) \\
&={\rm(transposition\ and\ 1\ sign\ change)}
\cdot\tl.
\end{array}
\end{displaymath}
This accounts for the $\e_{0,0}$ contributions in the odd case.

If $\l_\h=0$, then
\begin{displaymath}
\begin{array}{rl}
\tl-(\ldots,1)
&=(\ldots,\frac12)-(\ldots,1) \\
&=(\ldots,-\frac12) \\
&={\rm(1\ sign\ change)}
\cdot\tl.
\end{array}
\end{displaymath}
This accounts for the $\e_0$ contribution in the odd case.

If $\l_\h=1/2$, then
\begin{displaymath}
\begin{array}{rl}
\tl-(\ldots,2)
&=(\ldots,1)-(\ldots,2) \\
&=(\ldots,-1) \\
&={\rm(1\ sign\ change)}
\cdot\tl.
\end{array}
\end{displaymath}
This accounts for the $\e_{1/2}$ contribution in the odd case.

There are also $\e_0P^{(1)}$ and $\e_{1/2}P^{(1)}$ contributions in
the odd case; these arise as follows.  If $\l_\h=0$, $\l_a=\l_{a+1}\,$,
and $a+1<\h$, then
$$
\begin{array}{rl}
\tl-(1,-1,1)&
=(\ldots,\l_a\,,\l_a-1,\ldots,\frac12)-(1,-1,1) \\
&=(\ldots,\l_a-1,\l_a\,,\ldots,-\frac12) \\
&={\rm(transposition\ and\ 1\ sign\ change)}\cdot\tl.
\end{array}
$$
If $\l_\h=1/2$, $\l_a=\l_{a+1}\,$,
and $a+1<\h$, then
$$
\begin{array}{rl}
\tl-(1,-1,1)&
=(\ldots,\l_a\,,\l_a-1,\ldots,1)-(1,-1,2) \\
&=(\ldots,\l_a-1,\l_a\,,\ldots,-1) \\
&={\rm(transposition\ and\ 1\ sign\ change)}\cdot\tl.
\end{array}
$$
 
For all other values of $\tl':=\tl-\m$, where $\m\in\P(\l)$, either
$(|\tl'_1|,\ldots,|\tl'_\h|)$ is not a permutation of  
$(|\tl_1|,\ldots,|\tl_\h|)$,
or else $n$ is even and $\tl'_1\,\ldots\tl'_\h<0$.  (Recall that
we have assumed $\tl_\h\ge 0$.)  Thus the above exhausts all possible
contributions.

If $n=8$, we have the following changes to the case $n\ge 10$ even
just considered.  The weight $(1,1,1,1)$ is replaced by the two weights
$(1,1,1,\pm 1)$.  The entries 
$$
\begin{array}{l}
(1,1,1,-1)+\a_{ab}^-\sim(2,1,1)\qquad(3\ge b>a), \\
(1,1,1,-1)+\a_{a4}^+\sim(2,1,1)\qquad(3\ge a)
\end{array}
$$
need to be added to \nnn{RtStrE}.  A review of the weight multiplicity
calculation shows that each weight $(1,1,1,\pm 1)$ ``still'' has 
multiplicity 2.  Each weight $(1,1,1,\pm 1)$ has a Weyl group orbit of
size $8$.
(If we continue the formula for the size of the
Weyl orbit of $(1,1,1,1)$ in large even dimension
to dimension $8$, we get $16$; thus the orbit ``splits equally''
in the descent to dimension $8$.)
Reviewing the contributions to the Brauer-Kostant formula, the $S$
contribution is attributable to $(1,1,1,1)$, while the 
$\e_{0,0}P(\l^{(2)})$ contribution is attributable to $(1,1,1,-1)$.

When we descend to $n=6$, the weight $(1,1,1,1)$ disappears.
The weight $(2,1,1)$ ``splits'' into the weights $(2,1,\pm 1)$;
each has an orbit size of $12$, or half of that predicted by continuing
the expression $4\h(\h-1)(\h-2)$.  The first entry of \nnn{RtStrE}
is replaced by
$$
(2,1,\pm 1)+\a_{23}^\mp=(2,2),
$$
and so the multiplicity of each weight $(2,1,\pm 1)$ is $1$.
Note that the orbit size formula for this weight gives $0$ when $n=6$ is
substituted, and that the quantities $S$ and $\e_{0,0}P(\l^{(2)})$
vanish.  
The $T$ contributions, as well
as the $\e_{0,0,0}$ contributions, are evenly split, by
(\ref{TContrT},\ref{TContrH},\ref{EoooContr}).  

When $n=4$, there is only a single 
positive root string through dominant weights in each module
$V(2,\pm 2)$, namely
the $\a_{12}^\pm$ string starting at $0$.
The weights and multiplicities in the two tables result.
Note that $(2)$ is not a weight in either module, and that the formula for
its multiplicity as a weight in the large even $n$ case, namely $\h-2$,
evaluates to $0$ when $n=4$.  The quantities $S$, $T$ and $\e_{0,0,0}$ vanish
identically.  The $D$, $P$, $\e_{1/2,1/2}\,$, and $\e_{0,0}$ fall where
they do by (\ref{OneDrop},\ref{Pair},\ref{EHalfHalf},\ref{EooH}).

When $n=7$, 
the large odd $n$ case changes as follows.
The weight $(1,1,1,1)$ disappears; the formula for its orbit
size
gives the value $0$, and the quantity $S$ vanishes.
When $n=5$, the weights with $3$ nonzero entries also disappear, and
the expressions for their orbit sizes vanish. The quantities
$S$, $T$, and $P(\l^{(1)})$ vanish identically.

This completes the proof of Theorem \ref{WeightTables}. 
\end{Proof}

We can check some of our numbers by multiplying multiplicities by
orbit sizes and adding; this should give the dimension of the module
$V(2,2)$ (and $V(2,-2)$ when $n=4$).  Doing this for $n\ge 6$ even, we get
\begin{equation}\label{dimnE}
\frac{\h(\h+1)(2\h+1)(2\h-3)}{3}\,.
\end{equation}
This checks against direct computation via Weyl's dimension formula
\cite{Hu}.
When $n=4$ \nnn{dimnE} is still an expression for $\dim\cC$; and its
summands $V(2,\pm 2)$ are both $5$-dimensional.

When $n\ge 5$ is odd, the sum of multiplicities times orbit sizes is
\begin{equation}\label{dimnO}
\frac{(\h+1)(\h-1)(2\h+1)(2\h+3)}{3}\,.
\end{equation}
This checks against Weyl's dimension formula.
The quantities \nnn{dimnE} and \nnn{dimnO} have a unified expression in
terms of the dimension $n$, namely
$$
\frac{n(n+1)(n+2)(n-3)}{12}\,.
$$

It remains to add up the contributions to the Brauer-Kostant quantity.

\begin{Theorem}\label{ThmEven} 
Suppose $n\ge 4$ is even, and let
$$
\cC:=\left\{\begin{array}{ll}
V(2,2), & n\ge 6, \\
V(2,2)\os V(2,-2),\qquad & n=4.
\end{array}\right.
$$
If $\l_\h\ge 0$, then
\begin{equation}\label{EFormula}
\begin{array}{l}
\cM_\l(\cC\ot V(\l))=
\h(\h-1)+T-D+2S-(2\h-3)P \\
{}\qquad+\e_{0,0,0}
-\e_{1/2,1/2}+(P(\l^{(2)})-\h+1)\e_{0,0}\,.
\end{array}
\end{equation}
If $\l_\h<0$, then
$$
\cM_\l(\cC\ot V(\l))=\cM_{\bar\l}(\cC\ot V(\bar\l)),
$$
where $\bar\l=(\l_1\,,\ldots,\l_{\h-1}\,-\l_\h)$.
\end{Theorem}

\begin{Proof}{Proof} 
For $n\ge 8$, this is just a matter of adding up.
For $n=6$, we add up and recall the fact, noted above, that 
$S$ and $\e_{0,0}P(\l^{(2)})$
vanish identically.  
For $n=4$, we add up and recall that $S$, $T$ and $\e_{0,0,0}$ vanish
identically.

To handle the case in which $\l_\h<0$, we just need
to note the symmetry of the problem with respect to the reflection 
$\m\mapsto\bar\m$ on $\P$.
\end{Proof}

\begin{Theorem}\label{ThmFour} Suppose $n=4$.  If $\l_2\ge 0$, then
$$
\begin{array}{rl}
\cM_\l(V(2,2)\ot V(\l)) & =1
-\e_{1/2,1/2}-\e_{0,0}\,, \\
\cM_\l(V(2,-2)\ot V(\l)) & =1-D-P.
\end{array}
$$
If $\l_2<0$, then 
$$
\cM_\l(V(2,\pm 2)\ot V(\l))=\cM_{\bar\l}(V(2,\mp 2)\ot V(\bar\l)).
$$
\end{Theorem}

Theorem \ref{ThmFour} is a refinement of \ref{ThmEven} in the case 
$n=4$, since $\cC\ot V(\l)=V(2,2)\ot V(\l)\,\os\,V(2,-2)\ot V(\l)$.

\begin{Proof}
This is just a matter of adding up.
Again, to handle the case in which $\l_\h<0$, we just need
to note the symmetry of the problem with respect to the reflection 
$\m\mapsto\bar\m$ on $\P$.
\end{Proof}

\begin{Theorem}\label{ThmOdd} If $n\ge 5$ is odd, then
\begin{equation}\label{OFormula}
\begin{array}{l}
\cM_\l(V(2,2)\ot V(\l))=
(\h+1)(\h-1)+T-D+2S-2(\h-1)P \\
{}\qquad+(P(\l^{(1)})-\h+1)(2\e_0+\e_{1/2})+\e_{0,0}\,.
\end{array}
\end{equation}
\end{Theorem}

\begin{Proof}{Proof} This is just a matter of adding up.
Recall that when $n=7$,
the quantity $S$ vanishes identically, and that
when $n=5$, the quantities
$S$, $T$, and $P(\l^{(1)})$ vanish identically.
\end{Proof}

\section{Some more compact expressions}

Now suppose that $\l_\h\ge 0$, and write 
\begin{equation}\label{strings}
\l=(\underbrace{\a_1\,,\ldots,\a_1}_{k_1}\,,\ldots,
\underbrace{\a_r\,,\ldots,\a_r}_{k_r}),
\end{equation}
where $\a_1>\ldots>\a_r\,$.  In other words, group the strings of
identical entries in $\l$, and denote the string lengths by $k_1\,,\ldots,
k_r\,$.  Let $\lred$ be the result of eliminating singleton strings
(strings with $k_i=1$) from $\l$.  The tuple $\lred$ may have anywhere
from $0$ to $\h$ entries.
In analogy with \nnn{strings}, write
$$
\lred=(\underbrace{\b_1\,,\ldots,\b_1}_{x_1}\,,\ldots,
\underbrace{\b_s\,,\ldots,\b_s}_{x_s}).
$$
We claim that the terms in \nnn{EFormula} and \nnn{OFormula}
have very simple expressions in terms of $r$ and $s$.

To see this, first note that if
$$
X:=x_1+\ldots+x_s\,,
$$
then
\begin{equation}\label{preSimp}
\begin{array}{rl}
T&=X-2s, \\
2S&=\displaystyle
\sum_{i=1}^s(x_i-2)(x_i-3)+2\sum_{1\le i<j\le s}(x_i-1)(x_j-1), \\
&=X^2-(2s+3)X+s(s+5), \\
P&=X-s.
\end{array}
\end{equation}
Since
$$
r-s=\#({\rm singleton\ strings})=\h-X,
$$
this gives
\begin{displaymath}
\begin{array}{rl}
\h(\h-1)+T+2S-(2\h-3)P&=\left(X-s-\h+\tfrac12\right)^2+s-\tfrac14 \\
&=\left(\tfrac12-r\right)^2+s-\tfrac14 \\
&=r^2-r+s.
\end{array}
\end{displaymath}
Since 
$$
D=r-1+h,
$$
where 
$$
h:=\#\{a\mid\l_a\ge\l_{a+1}+2\},
$$
we have
\begin{equation}\label{simpT}
\h(\h-1)+T-D+2S-(2\h-3)P=(r-1)^2+h+s.
\end{equation}
($h$ may be described as the number of ``steep drops'' in $\l$.)

To simplify the block of terms
$$
\e_{0,0,0}+(P(\l^{(2)})-\h+1)\e_{0,0}\,,
$$
note that if $\e_{0,0,0}=1$, then
$$
\e_{0,0,0}+(P(\l^{(2)})-\h+1)\e_{0,0}=P(\l^{(2)})-\h+2.
$$
while if $\e_{0,0,0}=0$ and $\e_{0,0}=1$, then
$$
\e_{0,0,0}+(P(\l^{(2)})-\h+1)\e_{0,0}=P(\l^{(2)})-\h+1.
$$
In each case, this is the quantity
$$
P-\h=X-s-\h=-r.
$$
The conclusion is that 
\begin{equation}\label{simpH}
\e_{0,0,0}+(P(\l^{(2)})-\h+1)\e_{0,0}=-r\e_{0,0}\,.
\end{equation}
\nnn{simpT} and \nnn{simpH} give:

\begin{Theorem}\label{ThmEvenB}
Suppose $n\ge 4$ is even.  If $\l_\h\ge 0$, then
$$
\cM_\l(\cC\ot V(\l))=(r-1)^2+s+h-r\e_{0,0}-\e_{1/2,1/2}\,.
$$
If $\l_\h<0$, then $\cM_\l(\cC\ot V(\l))=\cM_{\bar\l}(\cC\ot V(\bar\l))$.
\end{Theorem}

If $n=4$, then $(r,s)$ is either $(2,0)$ or $(1,1)$, so that
$$
(r-1)^2+s=1.
$$
If $\e_{0,0}=1$, then $r=1$.  As a result, by Theorem \ref{ThmEvenB},
$$
\cM_\l(W\ot V(\l))=1+h-\e_{0,0}-\e_{1/2,1/2}\qquad(n=4,\ \l_2\ge 0).
$$
Together with the first equation of Theorem \ref{ThmFour}, this gives:

\begin{Theorem}\label{ThmFourB} Suppose $n=4$.  If $\l_2\ge 0$, then
$$
\begin{array}{rl}
\cM_\l(V(2,2)\ot V(\l)) & =1
-\e_{1/2,1/2}-\e_{0,0}\,, \\
\cM_\l(V(2,-2)\ot V(\l)) & =h.
\end{array}
$$
If $\l_2<0$, then 
$$
\cM_\l(V(2,\pm 2)\ot V(\l))=\cM_{\bar\l}(V(2,\mp 2)\ot V(\bar\l)).
$$
\end{Theorem}

Now consider the odd-dimensional case.  By \nnn{simpT} and \nnn{preSimp}, 
\begin{equation}\label{OddT}
\begin{array}{l}
(\h+1)(\h-1)+T-D+2S-2(\h-1)P=\h-1-P+(r-1)^2+h+s \\
{}\qquad=r(r-1)+h+s.
\end{array}
\end{equation}
If $\e_{0,0}=1$, then 
$$
2(P(\l^{(1)})-\h+1)\e_0+\e_{0,0}=2(P-\h)+1.
$$
If $\e_{0,0}=0$ but $\e_0=1$, then
$$
2(P(\l^{(1)})-\h+1)\e_0+\e_{0,0}=2(P-\h+1).
$$
Thus in all cases,
\begin{equation}\label{OddH}
2(P(\l^{(1)})-\h+1)\e_0+\e_{0,0}=2(P-\h+1)\e_0-\e_{0,0}=-2(r-1)\e_0
-\e_{0,0}\,.
\end{equation}
If $\e_{1/2,1/2}=1$, then
$$
(P(\l^{(1)})-\h+1)\e_{1/2}=P-\h.
$$
If $\e_{1/2,1/2}=0$ but $\e_{1/2}=1$, then
$$
(P(\l^{(1)})-\h+1)\e_{1/2}=P-\h+1.
$$
Thus in all cases,
\begin{equation}\label{OddE}
(P(\l^{(1)})-\h+1)\e_{1/2}=(P-\h+1)\e_{1/2}-\e_{1/2,1/2}
=-(r-1)\e_{1/2}-\e_{1/2,1/2}\,.
\end{equation}

Putting together \nnn{OddT}, \nnn{OddH}, and \nnn{OddE}, we have:

\begin{Theorem}\label{ThmOddB}
If $n\ge 5$ is odd, then
$$
\cM_\l(V(2,2)\ot V(\l))= 
r(r-1)+s+h-2(r-1)\e_0
-\e_{0,0}-(r-1)\e_{1/2}-\e_{1/2,1/2}\,.
$$
\end{Theorem}

Using the more compact formulas, it is possible to classify those
$\l$ for which $\cM_\l(V(2,2)\ot V(\l))=0$.  Suppose that
$n\ge 6$ is even, and $\l_\h\ge 0$.  If $r=1$, then $s=1$ and $h=0$, so
$$
\cM_\l(V(2,2)\ot V(\l))=1-\e_{0,0}-\e_{1/2,1/2}\,.
$$
If $r=2$, then $s\ge 1$, and
$$
(r-1)^2+s\ge 2,\qquad{\rm with\ equality}\ \iff\ s=1.
$$
If $r\ge 3$, then 
$$
\cM_\l(V(2,2)\ot V(\l))\ge r^2-3r+1\ge 1.
$$
Thus
$\cM_\l(V(2,2)\ot V(\l))$ vanishes if and only if
$\l$ is $0$, $(\frac12\,,\ldots,\frac12)$, or $(1)$.
Letting $\l_\h$ take any sign, we just need to add 
$(\frac12\,,\ldots,\frac12,-\frac12)$ to this list.

If $n=4$ and $\l_2\ge 0$, then $\cM_\l(V(2,2)\ot V(\l))$
vanishes if and only if $\l$ is $0$ or $(\tfrac12\,,\tfrac12)$.
$\cM_\l(V(2,-2)\ot V(\l))$ vanishes if and only if $\l_1-\l_2$
is $0$ or $1$.  If $\l_2<0$, we just reverse the roles of 
$(2,2)$ and $(2,-2)$ in these statements.  Note that in particular,
$V(1)$ is no longer on the list of modules which cannot be acted
upon by $V(2,2)$.

For odd $n$, it is sometimes useful to rewrite Theorem \ref{ThmOddB}
as
\begin{equation}\label{EvenBAlt}
\cM_\l(V(2,2)\ot V(\l))= 
(r-1)(r-2\e_0-\e_{1/2})
+s+h
-\e_{0,0}-\e_{1/2,1/2}\,.
\end{equation}
Suppose $n\ge 7$ is odd.  If $r=1$, then $s=1$ and $h=0$, so
that
$$
\cM_\l(V(2,2)\ot V(\l))=1-\e_{0,0}-\e_{1/2,1/2}\,.
$$
If $r=2$, then $s\ge 1$, and 
\nnn{EvenBAlt} shows that
$$
\cM_\l(V(2,2)\ot V(\l))\ge 0,
\qquad{\rm with\ equality}\ \iff\ (h=0\ {\rm and}\ \e_{0,0}=1).
$$
If $r\ge 3$, \nnn{EvenBAlt} shows that $\cM_\l(V(2,2)\ot V(\l))\ge 1$.
Thus 
$\cM_\l(V(2,2)\ot V(\l))$ vanishes if and only if
$\l$ is $0$, $(\frac12\,,\ldots,\frac12)$, or $(1)$.

If $n=5$, the discussion immediately above is altered as follows.
When $r=2$, one can no longer conclude that $s\ge 1$; in fact
$s$ is necessarily $0$.
\nnn{EvenBAlt} becomes
$$
\cM_\l(V(2,2)\ot V(\l))= 
2-2\e_0-\e_{1/2}
+h.
$$
This can vanish only for $\e_0=1$ and $h=0$.  But this is the situation only
for $\l=(1)$.
Thus the answer for $n=5$ is the same as for odd $n\ge 7$.

Summarizing, we have:

\begin{Theorem}\label{zeros}
If $n\ge 6$ is even, then  
$\cM_\l(V(2,2)\ot V(\l))$
vanishes if and only if $\l$ is $0$, 
$(\frac12\,,\ldots,\frac12\,,\pm\tfrac12)$, or $(1)$.
If $n=4$, then
$\cM_\l(V(2,2)\ot V(\l))$
vanishes if and only if $\l_1+\l_2\in\{0,1\}$, and
$\cM_\l(V(2,-2)\ot V(\l))$
vanishes if and only if $\l_1-\l_2\in\{0,1\}$.
If $n\ge 5$ is odd, then
$\cM_\l(V(2,2)\ot V(\l))$
vanishes if and only if $\l$ is $0$, 
$(\frac12\,,\ldots,\frac12)$, or $(1)$.
\end{Theorem}

It is also of some interest to know when there is a unique action of
the Weyl module; i.e., when 
$\cM_\l(V(2,2)\ot V(\l))$ (and $\cM_\l(V(2,-2)\ot V(\l))$
when $n=4$) is $1$.  We shall chase this through case by case,
omitting some of the (by now routine) details.

For $n\ge 6$ even, 
we have $\cM_\l(V(2,2)\ot V(\l))=1$ in the following cases:
\begin{equation}\label{OneAcE}
\begin{array}{ll}
r=1:\ &(p,\ldots,p,\pm p),\ {\rm where}\ p\ge 1; \\
r=2:\ &(\tfrac32\,,\tfrac12\,,\ldots,\tfrac12\,,\pm\tfrac12), \\
&(p,0,\ldots,0)\qquad
p\in 2+\bbN, \\
&(\underbrace{1,\ldots,1}_{{\rm at\ least}\ 2},
\underbrace{0,\ldots,0}_{{\rm at\ least}\ 2}); \\
r\ge 3:\ &{\rm none}.
\end{array}
\end{equation}

For $n=4$, $\cM_\l(V(2,2)\ot V(\l)$ and 
$\cM_\l(V(2,-2)\ot V(\l))$ can only take on the values $0$ and $1$;
thus the multiplicity $1$ case is exactly the complement of the
multiplicity $0$ case studied above.

If $n\ge 5$ is odd, we have
$\cM_\l(V(2,2)\ot V(\l))=1$ in the following cases:
\begin{equation}\label{OneAcO}
\begin{array}{ll}
r=1:\ &(p,\ldots,p),\ {\rm where}\ p\ge 1; \\
r=2:\ &(\tfrac32\,,\tfrac12\,,\ldots,\tfrac12), \\
&(1,\ldots,1,0)\qquad(n\ge 7), \\
&(p),\qquad
p\in 2+\bbN, \\
&(\underbrace{1,\ldots,1}_{{\rm at\ least}\ 2},
\underbrace{0,\ldots,0}_{{\rm at\ least}\ 2}); \\
r\ge 3:\ &{\rm none}.
\end{array}
\end{equation}

Summarizing, we have:

\begin{Theorem}\label{ones}
If $n\ge 6$ is even, then  
$\cM_\l(V(2,2)\ot V(\l))=1$
if and only if $\l$ is one of the dominant weights listed in 
\nnn{OneAcE}.  If $n=4$, then
$\cM_\l(V(2,2)\ot V(\l))=1$
if and only if $\l_1+\l_2\not\in\{0,1\}$,
$\cM_\l(V(2,-2)\ot V(\l))=1$
vanishes if and only if $\l_1-\l_2\not\in\{0,1\}$.
If $n\ge 5$ is odd, then  
$\cM_\l(V(2,2)\ot V(\l))=1$
if and only if $\l$ is one of the dominant weights listed in 
\nnn{OneAcO}.
\end{Theorem}

\section{Other actions of the Riemann curvature}

The Riemann curvature tensor $R$ is a section of the bundle
$$
\cR=\left\{\begin{array}{ll}
\cV(2,2)\os\cV(2)\os\cV(0),\qquad & n\ge 5, \\
\cV(2,2)\os\cV(2,-2)\os\cV(2)\os\cV(0),\qquad & n=4, \\
\cV(2)\os\cV(0),\qquad & n=3, \\
\cV(0),\qquad & n=2.
\end{array}\right.
$$
The Weyl tensor part(s), if any, lives in the bundle(s) $\cV(2,\pm 2)$.
The $\cV(2)$ part carries the Einstein (trace-free Ricci) tensor, and
the $\cV(0)$ part carries the scalar curvature.  (See \cite{Str} for
a detailed discussion.)  To count the actions of the ``rest'' of the
Riemann tensor, first note that $V(0)\ot V(\l)\cong_{\gso(n)}V(\l)$,
so there is always one action (namely multiplication) of the scalar curvature.
The computation of the number of actions of $V(2)$ on $V(\l)$ 
is much easier than that for the Weyl parts.  In fact, this computation
was incidental to an investigation of conformally invariant operators
in \cite{pspum}.
Let $n\ge 3$.  By \cite{Feg}, the direct sum decomposition of 
$V(1)\ot V(\l)$ into irreducibles is multiplicity-free, and
contains the module $V(\s)$ if and only if $\s$ is dominant, and either
$\s=\l\pm e_a$ for some $a$, or else $n$ is odd, $\l_\h>0$, and $\s=\l$.
(This computatation is accomplished quite easily using the Brauer-Kostant
formula.)  \cite{pspum} shows that if $N(\l)$ is the number of
irreducible summands in $V(1)\ot V(\l)$, then 
$$
\cM_\l(V(2)\ot V(\l))=\left[\frac{N(\l)-1}{2}\right]\,.
$$
But the discussion immediately above shows that $N(\l)$ is given,
in the notation of Sec.\ \ref{compusection} above, by
$$
N(\l)=\left\{\begin{array}{l}
2\h-2P-\e_{0,0},\qquad n\ge 4\ {\rm even},\ \l_\h\ge 0, \\
N(\bar\l),\qquad n\ge 4\ {\rm even},\ \l_\h<0, \\
2\h+1-2P-2\e_0-\e_{1/2}\,,\qquad n\ge 3\ {\rm odd}.
\end{array}\right.
$$
As a result, we have:

\begin{Theorem}\label{V2acs}
$$
\cM_\l(V(2)\ot V(\l))=\left\{\begin{array}{l}
\h-P-1=r-1,\qquad n\ge 4\ {\rm even},\ \l_\h\ge 0, \\
\cM_{\bar\l}(V(2)\ot V(\bar\l)),\qquad n\ge 4\ {\rm even},\ \l_\h<0, \\
\h-P-\e_0-\e_{1/2}=r-\e_0-\e_{1/2}
\,,\qquad n\ge 3\ {\rm odd}.
\end{array}\right.
$$
\end{Theorem}

In particular, we see immediately that the 
Einstein tensor cannot act on the bundle $\cV(\l)$ for precisely the following
dominant weights $\l$:
\begin{equation}\label{EinCantAct}
\begin{array}{ll}
(p,\ldots,p,\pm p),\qquad & n\ge 4\ {\rm even}; \\
0\ {\rm and}\ (\tfrac12\,,\ldots,\tfrac12),\qquad & n\ge 3\ {\rm odd}.
\end{array}
\end{equation}

We can also sum up the contributions of the Weyl, Einstein, and scalar 
curvatures to get:

\begin{Theorem}\label{RiemAc} 
$$
\cM_\l(\cR\ot V(\l))=\left\{\begin{array}{ll}
r(r-1)+s+h-r\e_{0,0}-\e_{1/2,1/2}\,,\ &n\ge 4\ {\rm even},\ \l_\h\ge 0, \\
\cM_{\bar\l}(\cR\ot V(\bar\l)),\ &n\ge 4\ {\rm even},\ \l_\h<0, \\
r^2+s+h+1-(2r-1)\e_0-\e_{0,0} \\
{}\qquad-r\e_{1/2}-\e_{1/2,1/2}\,,\ &n\ge 3\ 
{\rm odd}.
\end{array}\right.
$$
\end{Theorem}

Because of the scalar curvature action, $\cM_\l(\cR\ot V(\l))$ is always
at least $1$.  It is {\em exactly} $1$ in precisely the following cases:
$$
\begin{array}{ll}
0\ {\rm or}\ (\tfrac12\,,\ldots,\tfrac12\,,\pm\tfrac12),\qquad & n\ge 4\ 
{\rm even}, \\
0\ {\rm or}\ (\tfrac12\,,\ldots,\tfrac12),\qquad & n\ge 3\ {\rm odd}.
\end{array}
$$ 

\section{Some explicit tensor representations}
The modules which figure prominently in the discussion immediately
above all
have more or less explicit tensor, or tensor-spinor, realizations.
For $n$ odd, the {\em spinor} module $\S$ 
is irreducible, and has highest weight 
$(\tfrac12\,,\ldots,\tfrac12)$.  If $n$ is even, the spinor module
splits as $\S_+\os\S_-\cong_{\gso(n)}
V(\tfrac12\,,\ldots,\tfrac12)\os V(\tfrac12\,,\ldots,\tfrac12
-\tfrac12)$.  If $k<n/2$, the module
$$
V(\underbrace{1,\ldots,1}_k,0,\ldots,0)
$$
is realized by the alternating $k$-tensors $\L^k$, and also by
$\L^{n-k}$.  (The {\em Hodge star} operator provides an explicit
equivariant isomorphism between these two realizations.)  If $n$ is
even, the modules
$V(1,\ldots,1,\pm 1)$ are realized in middle forms of the two different
dualities (eigenvalues under the Hodge star), $\L^{n/2}_\pm\,$.
If $p$ is a natural number, the module $V(p)$ is realizable as that of
trace-free symmetric $p$-tensors.  The Weyl module $\cC$ may be realized
as that of totally trace-free $4$-tensors $C_{abcd}$ with 
$$
C_{abcd}=C_{cdab}=-C_{bacd}=-C_{acdb}-C_{adbc}\,.
$$
If $n$ is odd, the module $V(\tfrac32\,,\tfrac12\,,\ldots,\tfrac12)$
may be realized as the {\em twistors}; i.e.\ spinor-one-forms
$\f_a$ with $\g^a\f_a=0$.  Here $\g^a$ are the Clifford matrices,
only tensor indices are written, and summation convention is understood.
If $n$ is even, the twistors split as
$V(\tfrac32\,,\tfrac12\,,\ldots,\tfrac12)\os V(\tfrac32\,,\tfrac12\,,
\ldots,\tfrac12\,,-\tfrac12)$.  More generally, modules of the
form 
$$
V(\tfrac32\,,\ldots,\tfrac32\,,\tfrac12\,,
\ldots,\tfrac12\,,\pm\tfrac12)\ {\rm or}\ 
V(\tfrac32\,,\ldots,\tfrac32\,,
\pm\tfrac32)
$$
are direct summands of modules of spinor-forms (see \cite{tbadv},
\cite{soubook}).
A module of the form $V(p,\ldots,p,\pm p)$, for integral $p$, is a direct
summand (in fact, the highest weight summand) 
of the $p$-fold symmetric tensor power $S^p_\pm$ of $V(1,\ldots,1,\pm 1)$
(which in turn, by the above, is a differential form module).  For 
properly half-integral $p$, we get the direct summand of 
$\S_\pm\ot S^{p-\frac12}_\pm\,$ containing the highest weight vector.

The fact that $V(2)$ cannot act on $V(0)$ nor on
$V(\tf\,,\ldots,\tf\,,\pm\tf)$ is
actually immediate by weight considerations.  First note that
$$
\hom_{\gso(n)}(V(\s)\ot V(\l),V(\t))\cong
\hom_{\gso(n)}(V(\s),V(\l)^*V(\t)),
$$
where $V(\l)^*$ is the module dual to $V(\l)$.  It is easily seen that
$V(\l)^*\cong V(\l)$ unless $n$ has the form $4k+2$ and $\l_\h\ne 0$, in which
case $V(\l)^*\cong V(\bar\l)$.  (Since the dual is irreducible, we just have
to find the dominant weight in the Weyl group orbit of $-\l$.  This is
either $\l$ or $\bar\l$, as indicated.)  Denote the highest weight
of the dual module by $V(\l^*):=V(\l)^*$.  Then
$$
\l^*+\t<\s\Rightarrow\cM_\t(V(\s)\ot V(\l))=0,
$$
where the ``$<$'' relation on the left is the lexicographical ordering.
Since $0+0<2$ and $\frac12+\frac12<2$, 
there is no action of $V(2)$ on $V(0)$, nor on
$V(\tf\,,\ldots,\tf\,,\pm\tf)$.

For the same reason, $V(2,2)$ cannot act on the trivial module, nor
on any spinor bundle, nor on $\L^1$.  This immediately
gives the ``if'' half of Theorem \ref{zeros} for $n>4$.

These weight size considerations do not, however, give the complete
lists of modules that cannot be acted upon (see the $n=4$ case of 
Theorem \ref{zeros} and the first line of \nnn{EinCantAct}).
And, of course, weight size considerations say nothing about the
``only if'' part of these statements.

An elementary point of contact of these results with some results in
geometric analysis is visible when we look at the {\em Weitzenb\"ockian},
a curvature action on differential forms which is the difference
between the form Laplacian $\D$ and the Bochner Laplacian $\N^*\N$.
By \cite{gold}, p.\ 118, application of the Weitzenb\"ockian
to the $k$-form $\f_{a_1\,\ldots a_k}$ for $(k\ge 1)$ gives
\begin{equation}\label{goldberg}
\begin{array}{ll}
&(B\f)_{a_1\,\ldots a_k}=
\dfrac{n-2k}{n-2}r_{b[a_1}\f^b{}_{a_2\,\ldots a_k]} \\
&\qquad+\dfrac{k-1}{(n-1)(n-2)}\,K\f_{a_1\,\ldots a_k}
+\dfrac{k-1}2C_{bc[a_1\,a_2}\f^{bc}{}_{a_3\,\ldots a_k]}\,,
\end{array}
\end{equation}
where $K$, $r$, and 
$C$ are respectively the scalar, Ricci, and Weyl curvatures,
and indices within square brackets are to be skewed (antisymmetrized)
over.
(The reference actually gives a formula for $(B\f,\f)$, where 
$(\cdot,\cdot)$ is the natural inner product.  Since $B$ is symmetric,
the formula for $B\f$ can be recovered.)
If $k=n/2$, the $r$ coefficient vanishes, showing that the Einstein
tensor is not involved in the formula.  Our results show that it {\em cannot}
be involved, since (1) the Weitzenb\"ockian, like $\D$ and $\N^*\N$,
carries each of the two middle-form bundles $\L^{n/2}_\pm$ to itself;
(2) our result \nnn{EinCantAct} shows that there is no action of 
$V(2)$ on $V(1,\ldots,1,\pm 1)$.  The action of 
$V(2)$ on $\L^k$ implicitly exhibited in \nnn{goldberg}, namely
$$
\f_{a_1\,\ldots a_k}\mapsto
\s_{b[a_1}\f^b{}_{a_2\,\ldots a_k]}\,,\qquad\s\in V(2),
$$
when used on middle-forms,
thus has to interchange
the modules $\L^{n/2}_\pm\,$.

The absence of the Einstein
tensor in the middle-form
Weitzenb\"ockian is an important point in the derivation of the
Bourguignon Vanishing Theorem \cite{Bour}.
Similarly, the Weyl tensor part of the Weitzenb\"ockian must vanish
for $k=1$.

Among other things, the formula \nnn{goldberg} exhibits the action
of the Weyl module on $\L^k$ for $k\ne 0,1,n-1,n$.
(In the cases $k=n-1,n$, the expression vanishes by an argument
of ``skewing on too many indices.'')

The results of \cite{pspum}
show that each bundle $\cV(\l)$ admits either a $1-$ or a $0-$ dimensional
space of second-order conformally covariant differential operators, modulo
order $0$ actions of the Weyl tensor.  The results of this paper show
how many of these actions of the Weyl tensor there are -- in particular,
they show that generically, there are many.

Let $A$ be a scalar-valued function on the set $\P^{\rm DI}$
of dominant integral $\gso(n)$ weights.  We say that
$A$ {\em vanishes generically} if 
there is a finite set $F$ such
that
$$
\l_1\,\ldots,\l_\h\,,\l_1-\l_2\,,\l_2-\l_3\,,\ldots,\l_{\h-1}-|\l_\h|
\not\in F\;\Rightarrow\;a(\l)=0.
$$
If $A-B$ vanishes generically, we say that $A\stackrel{{\rm gen}}{=}B$.
For example,
$$
r\gene\h,\ s\gene 0, h\gene\h-1, 
$$
and
$$
\e_{\k_{\h-k}\,,\ldots,\k_\h}\gene 0
$$
for $k\ge 1$.  Our results say that $\cM_\l(V(2,2)\ot V(\l))\gene(\h+1)(\h-1)$
for $n\ge 5$ odd, and that $\cM_\l(\cC\ot V(\l))\gene\h(\h-1)$
for $n\ge 4$ even.  When $n=4$, 
$\cM_\l(V(2,2)\ot V(\l))\gene\cM_\l(V(2,-2)\ot V(\l))\gene 1$.

Theorem \ref{V2acs} indicates that there will be only one action 
of trace-free symmetric $2$-tensors
on trace-free symmetric $p$-tensors for $p\ge 1$; this is given by
$$
\f_{a_1\,\ldots a_p}\mapsto\s_{b(a_1}\f^b{}_{a_2\,\ldots a_p)_0}\,,
$$
where $(\cdots)_0$ is trace-free symmetrization on the enclosed indices.
Theorem \ref{V2acs} also says that a bundle of 
twistors, i.e. spinor-one-forms
$\y_a$ with $\g^a\y_a=0$, admits only one action of $V(2)$.  
(Recall that the twistor bundle is $V(\tfrac32\,,\tfrac12\,,\ldots,
\tfrac12)$ if $n\ge 3$ is odd, and 
$V(\tfrac32\,,\tfrac12\,,\ldots,
\tfrac12)\os V(\tfrac32\,,\tfrac12\,,\ldots,
\tfrac12,-\tfrac12)$ if $n\ge 4$ is even.)  There are two linearly
independent spinor one-forms that can be constructed from $\s\ot\y$, namely
$$
\s_a{}^b\y_b\qquad{\rm and}\qquad\s_b{}^c\g^b\g_a\y_c\,.
$$
But by the Clifford relation
$$
\g^a\g^b+\g^b\g^a=-2g^{ab},
$$
only one linear combination of these is a twistor, namely
$$
\s_a{}^b\y_b-\dfrac1{n-2}\s_b{}^c\g^b\g_a\y_c\,.
$$
Similar statements can be made, for $n\ge 6$, 
about bundles of spinor $2$-forms
with $\g^a\y_{ab}=0$; here the action is
$$
\s^c{}_{[a}\y_{b]c}-\dfrac1{n-4}\s_{cd}\g^c\g_{[a}\y_{b]d}\,.
$$
The highest weights of these bundles
consist are $(\frac32\,,\frac32\,,\frac12\,,\ldots,\frac12)$, and,
if $n$ is even, $(\frac32\,,\frac32\,,\frac12\,,\ldots,\frac12\,,-\frac12)$.
The generalization to spinor-$k$-forms $\y_{a_1\,\ldots a_k}$ with
$\g^{a_1}\y_{a_1\,\ldots a_k}$ is clear.

We get an action of the Weyl bundle on trace-free symmetric $p$-tensors
by taking
\begin{equation}\label{tfsC}
C^b{}_{(a_1}{}^c{}_{a_2}\f_{a_3\,\ldots a_p)_0bc}\,,
\end{equation}
as long as $p\ge 2$.  Theorems \ref{ThmEvenB} and \ref{ThmOddB}
predict that there should be just one action of the Weyl tensor
as long as $n\ge 5$.  If $n=4$, then by Theorem \ref{ThmFourB}, there
should be one action by $V(2,2)$ (the self-dual Weyl tensors), and
and one by $V(2,-2)$ (the anti-self-dual Weyl tensors).
Both of these are also described by \nnn{tfsC}.

By the Clifford relations and the fact that Weyl tensors are trace free,
$$
\f_a\mapsto C_{bc}{}^d{}_a\g^b\g^c\y_d
$$ 
is an action of the Weyl bundle on twistors; by Theorems 
\ref{ThmEvenB} and \ref{ThmOddB}, this is the only action for $n\ge 4$
which, in the even-dimensional case, preserves each subbundle
$(\frac32\,,\frac12\,,\ldots,\frac12\,,\pm\frac12)$.

When $n=4$, $V(2,\pm 2)$ is a submodule of $V(1,\pm 1)\ot V(1,\pm 1)$;
thus $\cC$ is a submodule of $\L^2_+\ot\L^2_+\,\os\,\L^2_-\ot\L^2_-\,$.
In particular, the ``mixed'' summands $\L^2_+\ot\L^2_-$ and
$\L^2_-\ot\L^2_+$ in $\L^2\ot\L^2$
do not contribute to $\cC$.  As a result, $V(2,\pm 2)$
should not be able to act on $V(1,\mp 1)$; this is confirmed by Theorem
\ref{ThmFourB}.  

\section{Epilogue}

A question which is not addressed in this paper, but would be well
worth the effort, is to compute the dimension of
$\hom_{\gso(n)}(\cC\ot V(\l), V(\k))$ for $\k\ne\l$.  The
corresponding problem with $V(2)$ in place of $\cC$ was addressed in
\cite{socc}, where it plays a part in classifying second-order
conformally invariant operators between different $\spin(n)$-bundles.

Whenever $V(1)\ot V(\l)$ has $V(\t)$ as a summand, there is a first-order
equivariant differential operator from $\cV(\l)$ to $\cV(\t)$, the so-called
{\em generalized gradient}.  This comes from compressing the covariant
derivative to act between irreducible summands:
$$
\cV(\l)\stackrel{\N}\rightarrow T^*M\ot\cV(\l)=
\cV(\t_1)\os\ldots\cV(\t_{N(\l)})\stackrel{{\rm Proj}_{\t_u}}{\rightarrow}
\cV(\t_u).
$$
(Recall that the decomposition $\cV(\t_1)\os\ldots\os\cV(\t_{N(\l)})$
is multiplicity free.)
The generalized gradient is the composition $G_{\t_u\l}={\rm Proj}_{\t_u}
\circe\N$.
When we compose two gradients, say
\begin{equation}\label{composition}
V(\l)\stackrel{G_1}{\leftarrow}V(\t)\stackrel{G_2}{\leftarrow}V(\k),
\end{equation}
we either get an operator of order $2$,
or (by Weyl's invariant theory), one of order $0$.  If 
$\hom_{\gso(n)}(V(2)\ot V(\l), V(\k))=0$, then we must get
an order $0$ operator, since the leading symbol of an order $2$
operator is canonically associated to an element of 
$\hom_{\gso(n)}(V(2)\ot V(\l), V(\k))$.  This order $0$ operator
is an action of the Riemann curvature, to which the scalar curvature
cannot contribute, since $\k\ne\l$.  The Einstein tensor cannot contribute
either, since it is a section of $V(2)$.  Thus in the situation outlined
here, the composition $G_2G_1$ is an action of the Weyl tensor.
Knowing how many actions of the Weyl tensor there are from $V(\l)$ to
$V(\k)$ might be valuable in the computatation of this action.
In particular, if there are no such actions, the composition must be
vanish.  This, in fact, is exactly what happens for the operators
$dd$ of the de Rham complex.

It is not too hard to show (see \cite{socc}) that the possible compositions
reaching $V(\k)$ from $V(\l)$, when $\k\ne\l$, are limited as follows.
In \nnn{composition}, there are either $1$ or $2$ choices for
$\t$.
When there is one choice, $G_2\circe G_1$ might have order $2$, or
might be an
action of the Weyl tensor.
When there are two choices, we have a diagram of operators
$$
\begin{array}{ccc}
& \tilde G_2 & \\
V(\tilde\t) & \quitelongrightarrow & V(\k) \\
&& \\
\uparrow\ \tilde G_1 & & \uparrow G_2 \\
&& \\
V(\l) & \quitelongrightarrow & V(\t) \\
& G_1 &
\end{array}
$$
and there is a unique (up to constant multiples) linear combination
of $G_2G_1$ and $\tilde G_2\tilde G_1$ which is an order
0 action of the Riemann curvature.
This may contain contributions from the Weyl tensor and the Einstein
(trace-free Ricci) tensor, but not from the scalar curvature.

Such considerations are behind important theorems on overdetermined
systems of differential equations on spinors \cite{berliners}, the 
solvability of which characterizes special manifolds.  It is hoped
that the present program of enumerating and classifying curvature
actions might eventually be used to get analogous results based on
twistors, higher spinor-forms, and sections of other geometric bundles.

\vspace{0.4in}

\noindent Collin Bennett, Department of Mathematical and Computer Sciences,
Loyola University, Chicago IL 60626 USA\newline
cbennet@math.luc.edu

\vspace{0.2in}

\noindent Thomas Branson, 
Department of Mathematics, University of Iowa, Iowa City IA 52242 USA\newline
branson@math.uiowa.edu

\end{document}